\documentclass{aa}
\usepackage{graphics,epsfig,amsmath,amssymb,amstext,txfonts}

\def\la{\mathrel{\mathchoice {\vcenter{\offinterlineskip\halign{\hfil
$\displaystyle##$\hfil\cr<\cr\sim\cr}}}
{\vcenter{\offinterlineskip\halign{\hfil$\textstyle##$\hfil\cr
<\cr\sim\cr}}}
{\vcenter{\offinterlineskip\halign{\hfil$\scriptstyle##$\hfil\cr
<\cr\sim\cr}}}
{\vcenter{\offinterlineskip\halign{\hfil$\scriptscriptstyle##$\hfil\cr
<\cr\sim\cr}}}}}

\def\ga{\mathrel{\mathchoice {\vcenter{\offinterlineskip\halign{\hfil
$\displaystyle##$\hfil\cr>\cr\sim\cr}}}
{\vcenter{\offinterlineskip\halign{\hfil$\textstyle##$\hfil\cr
>\cr\sim\cr}}}
{\vcenter{\offinterlineskip\halign{\hfil$\scriptstyle##$\hfil\cr
>\cr\sim\cr}}}
{\vcenter{\offinterlineskip\halign{\hfil$\scriptscriptstyle##$\hfil\cr
>\cr\sim\cr}}}}}

\begin{document}

\title{UHE nuclei propagation and the interpretation of the ankle in the cosmic-ray spectrum}

\author{D. Allard\inst{1,2,3} \and E. Parizot\inst{1} \and E. Khan\inst{1} \and S. Goriely\inst{4} \and A. V. Olinto\inst{2,3}}

\institute{Institut de Physique Nucl\'eaire d'Orsay, IN2P3-CNRS/Universit\'e Paris-Sud, 91406 Orsay Cedex, France \and Kavli Institute of Cosmological Physics, University of Chicago, 5640 S. Ellis, Chicago, IL60637, USA \and Department of Astronomy and Astrophysics, University of Chicago, 5640 S. Ellis, Chicago, IL60637, USA \and Institut d'Astronomie et d'Astrophysique, ULB - CP226, 1050 Brussels, Belgium}

\offprints{parizot@ipno.in2p3.fr}

\date{Received date; accepted date}

\abstract{We consider the stochastic propagation of high-energy protons and nuclei in the cosmological microwave and infrared backgrounds, using revised photonuclear cross-sections and following primary and secondary nuclei in the full 2D nuclear chart. We confirm earlier results showing that the high-energy data can be fit with a pure proton extragalactic cosmic ray (EGCR) component if the source spectrum is $\propto E^{-2.6}$. In this case the ankle in the CR spectrum may be interpreted as a pair-production dip associated with the propagation. We show that when heavier nuclei are included in the source with a composition similar to that of Galactic cosmic-rays (GCRs), the pair-production dip is not present unless the proton fraction is higher than 85\%. In the mixed composition case, the ankle recovers the past interpretation as the transition from GCRs to EGCRs and the highest energy data can be explained by a harder source spectrum $\propto E^{-2.2}$-- $E^{-2.3}$, reminiscent of relativistic shock acceleration predictions, and in good agreement with the GCR data at low-energy and holistic scenarios.

}

\authorrunning{D. Allard, E. Parizot, E. Khan, S. Goriely and A. V. Olinto}
\titlerunning{UHE nuclei propagation and the interpretation of the CR ankle}

\maketitle

\section{Introduction}
\label{sec:introduction}

One of the keys to understanding the origin of cosmic-rays (CR) is the spectral shape of the transition from Galactic cosmic-rays (GCR) to extragalactic cosmic rays (EGCR). The Galactic origin of low-energy CRs is generally accepted, while the highest energy CRs  (above $\sim 10^{19}$~eV) are no longer confined by Galactic magnetic fields and most probably originate from other galaxies (see, however, Dar and Plaga, 1999; Blasi et al., 2000; Plaga, 2002). Thus, a transition between the two components has to occur in some energy range. The most natural location for this transition is around $\sim 3\,10^{18}$~eV at a feature in the CR spectrum known as the \emph{ankle}. This is the only energy range where the spectrum gets harder (i.e., its logarithmic slope gets smaller), offering a simple transition scheme to a harder CR component which is subdominant at lower energies.

A different conclusion has recently been proposed on the basis of composition results from the HiRes Collaboration, tentatively showing a transition from heavy to light primary nuclei at an energy around $5\,10^{17}$~eV (Abbasi et al. 2005), which can be identified with a \emph{second knee} feature in the spectrum. Such composition measurements rely on statistical determinations of the elongation rate of extensive air-showers and depend on shower development simulations which are still quite uncertain. One should thus remain cautious about such a result, but taken at face value it suggests a transition from a heavy GCR component to a light EGCR one at an energy \emph{below} the ankle.  
A  phenomenological class of models  where the GCR/EGCR transition is around the second knee can be designed such that the extragalactic component accounts for CRs down to energies $\la 10^{18}$~eV (Berezinsky et al., 2004). In these models, the ankle is interpreted as an  e$^{+}$-e$^{-}$ pair-production dip resulting from proton propagation over large distances in the cosmic microwave background (CMB). This effect is the pair production analog of the GZK flux suppression due to photo-pion production at higher proton energies (Greisen, 1966; Zatsepin and Kuzmin, 1966).
As in any model where the spectrum changes from a harder to a softer component, this class of models requires some unavoidable ``fine tuning''. Part of the fine tuning in the region of a GCR/EGCR transition can be explained if the extragalactic magnetic field has a high enough (but reasonable) intensity to prevent lower energy CRs from reaching us from distant galaxies (Lemoine, 2004; Aloiso and Berezinsky, 2004; see also Parizot, 2004, for a comparison of the critical magnetic field with an equipartition value). While the composition measurements and the matching of GCRs to EGCRs at the second knee remain unclear, the main point to note is that the transition \emph{does not have} to be at the ankle.

Phenomenological approaches are useful since none of the GCR and EGCR components are well understood yet, and their origin remains unknown (see, e.g., Parizot, 2005). Concerning the EGCR component, although major uncertainties remain in the ultra-high-energy (UHE) range, where the expected GZK suppression of the flux may not be found in the data (e.g., the AGASA spectrum in Teshima et al., 2003), propagation studies find that the measured spectrum is best fitted by assuming a power-law source spectrum in $E^{-\beta}$ with $\beta \simeq 2.6$ (De Marco et al., 2003), as also found by the second-knee-transition models. These studies, however, assume that EGCRs are only protons.

In this \emph{Letter}, we investigate how these conclusions are modified when nuclei heavier than protons are also present in the source. We apply the revised scheme of photo-nuclear interactions described in Khan et al. (2005). Interestingly, we find that the two main aspects of EGCR phenomenology are jointly affected, namely the best fit slope of the source spectrum and the interpretation of the ankle. The interpretation of the ankle as a pair production dip is not viable in the mixed composition case, unless the fraction of protons is above 85\% and spectral indices are fine tuned to ad hoc values. When reasonable assumptions for the fraction of nuclei in the EGCR are introduced, the traditional interpretation of a GCR/EGCR transition at the ankle is recovered and the best fit spectral index for EGCR is  $\sim$ 2.2 -- 2.3, as expected from relativistic shock acceleration. Finally, the best fit EGCR spectrum occurs when both the GCR and EGCR component have the same spectral index, allowing for  holistic approaches to the CR spectrum.

\section{Phenomenological source model}

Although extragalactic magnetic fields could modify the UHECR spectrum (e.g., Medina Tanco, 2001; Isola et al. 2002; Deligny et al., 2004), we assume here that their intensity is low enough to be negligible and focus on a comparison between pure proton and mixed-composition models. The choice of a source composition is arbitrary in the absence of a source model, but we consider a ``generic composition'' assuming that EGCRs have the same relative source abundances as the best known, low-energy GCRs, derived from \emph{Ulysses} and \emph{HEAO 3} data (Duvernois and Thayer, 1996). We convert the differential abundance ratios, $x_{i}$, at a given energy per nucleon ($E/A$) into ratios at a given energy, $\xi_{i} = x_{i}A_{i}^{\alpha-1}$, where $\alpha$ is the spectral index, so that the GCR source spectrum is: $N_{i}(E) \propto \xi_{i} E^{-\alpha}$. 

Another ingredient of our source model is the spectral index at high energy, $\beta$, which may be different from $\alpha$, raising the difficulty of whether we should use $x_{i}A_{i}^{\alpha-1}E^{-\beta}$ or $x_{i}A_{i}^{\beta-1}E^{-\beta}$ at the source. Since the CR composition depends essentially on the injection mechanism and the latter is uncertain for both EGCRs and GCRs, there is no reason, in principle, to expect that both components have the same differential composition. On the other hand, if EGCRs are accelerated out of the interstellar medium by a similar type of plasma process as GCRs, it is not unreasonable to assume a similar composition as well. 
Since our main goal is to investigate the difference between pure proton and mixed CR compositions, we study the general case of $x_{i}A_{i}^{\alpha-1}E^{-\beta}$ and $\alpha=\beta$ as a special case. Although we explore $\beta$ between 1 and 2.7, physically motived spectral indices are typically $ \beta \sim $2.2 -- 2.3 for the source spectrum. 

Finally, we choose a rigidity dependent maximum energy at the source for the various nuclear species  $E_{\mathrm{max}}(^{A}_{Z}\mathrm{X})$. We assume that energy losses and photo-fragmentation inside the source can be neglected, so that all nuclei with the same gyroradius behave the same way. This implies $E_{\mathrm{max}}(^{A}_{Z}\mathrm{X}) = Z\times E_{\mathrm{max}}(^{1}_{1}\mathrm{H})$.

\section{Propagation model}

We compute the propagated spectrum of EGCRs for a uniform source distribution with negligible magnetic fields, using a Monte-Carlo technique. For protons, we take into account the energy losses due to the photo-production of pions and e$^{+}$-e$^{-}$ pairs, caused by the interaction with CMB photons. Photo-pair production is treated as a continuous process, as allowed by the short interaction length and small inelasticity, while pion production is treated stochastically: we first draw randomly the energy of the interacting photon in the proton rest frame, from the relevant boosted black-body distribution weighted by the interaction cross-section, and then calculate the inelasticity from the reaction kinematics after choosing randomly the interaction angle in the center-of-mass frame. We also take into account the CMB temperature change as a function of time, as well as the energy losses due to the expansion of the universe.

The propagation of nuclei is also followed stochastically. In addition to e$^{+}$-e$^{-}$ pair production, nuclei are subject to photo-erosion, i.e., they lose nucleons through photo-nuclear interactions with the CMB and infrared background (IRB).  We use here for the first time a 2D scheme to follow the nuclei in the $(A,Z)$ space, thanks to both revised and new cross-sections as described in Khan et al. (2005). This includes the following processes: i) the giant dipolar resonance (GDR), with a loss of one or more nucleons, as well as $\alpha$-particles (this process occurs with photons above a threshold of $\sim 8$~MeV in the nucleus rest frame), ii) the quasi-deuteron (QD) process, where a virtual pion interacts with a nucleon pair within the nucleus, leading to the ejection of the pair and possibly additional protons or neutrons (this occurs with photons of typically 20~MeV), iii) baryonic resonances (BR), where a real pion is produced, ejecting a nucleon and possibly interacting further with a nucleon pair, eventually leading to the loss of an average of six nucleons for an Fe nucleus (this occurs above $\sim 150$~MeV), and iv) the photo-fragmentation (PF), occurring at very high energy ($\sim 1$~GeV) and breaking the nucleus into many fragments of much lower mass and energy.

\begin{figure*}[ht]
\centering
\hfill\includegraphics[height=5.5cm]{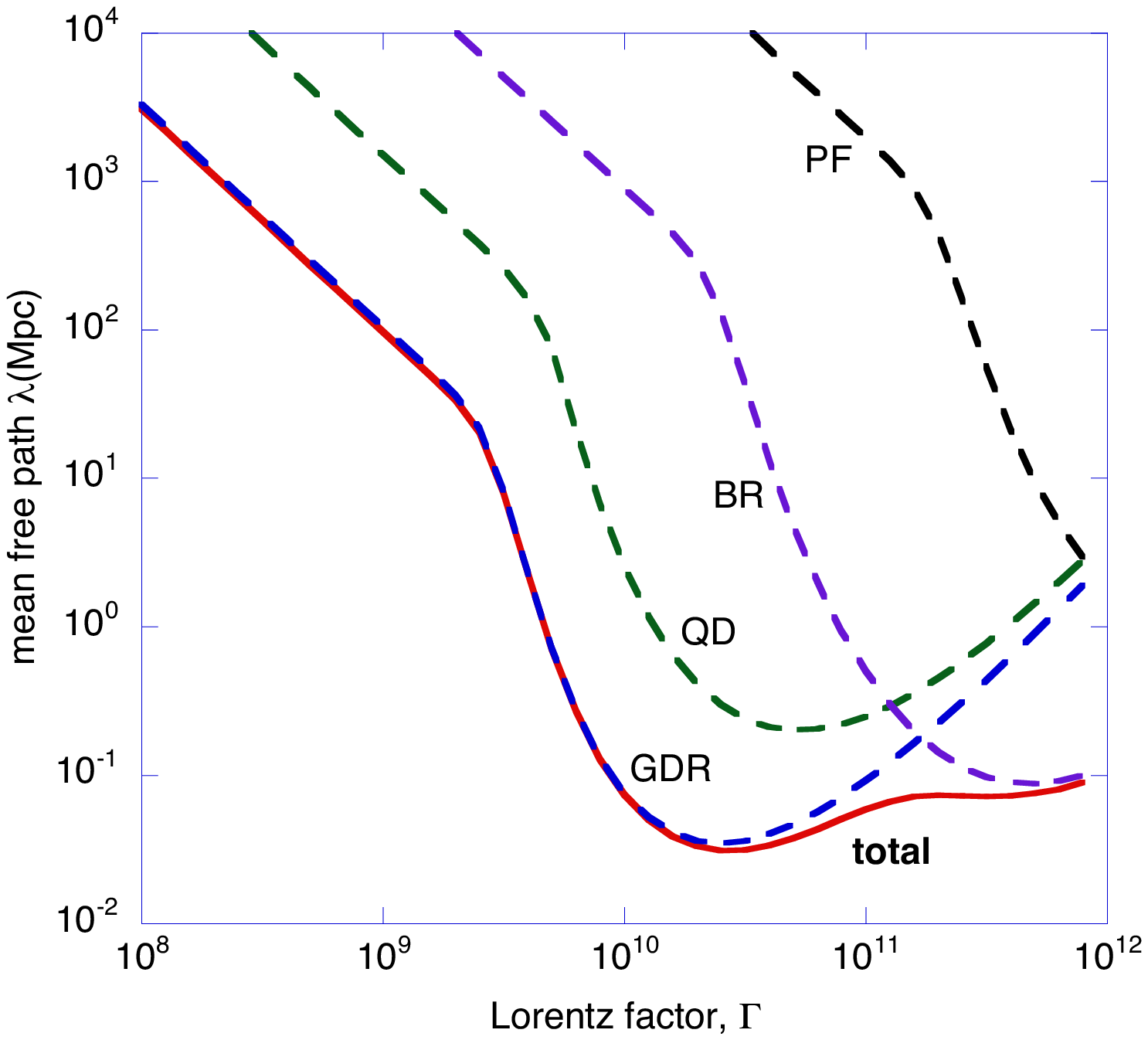}\hfill
\includegraphics[height=5.5cm]{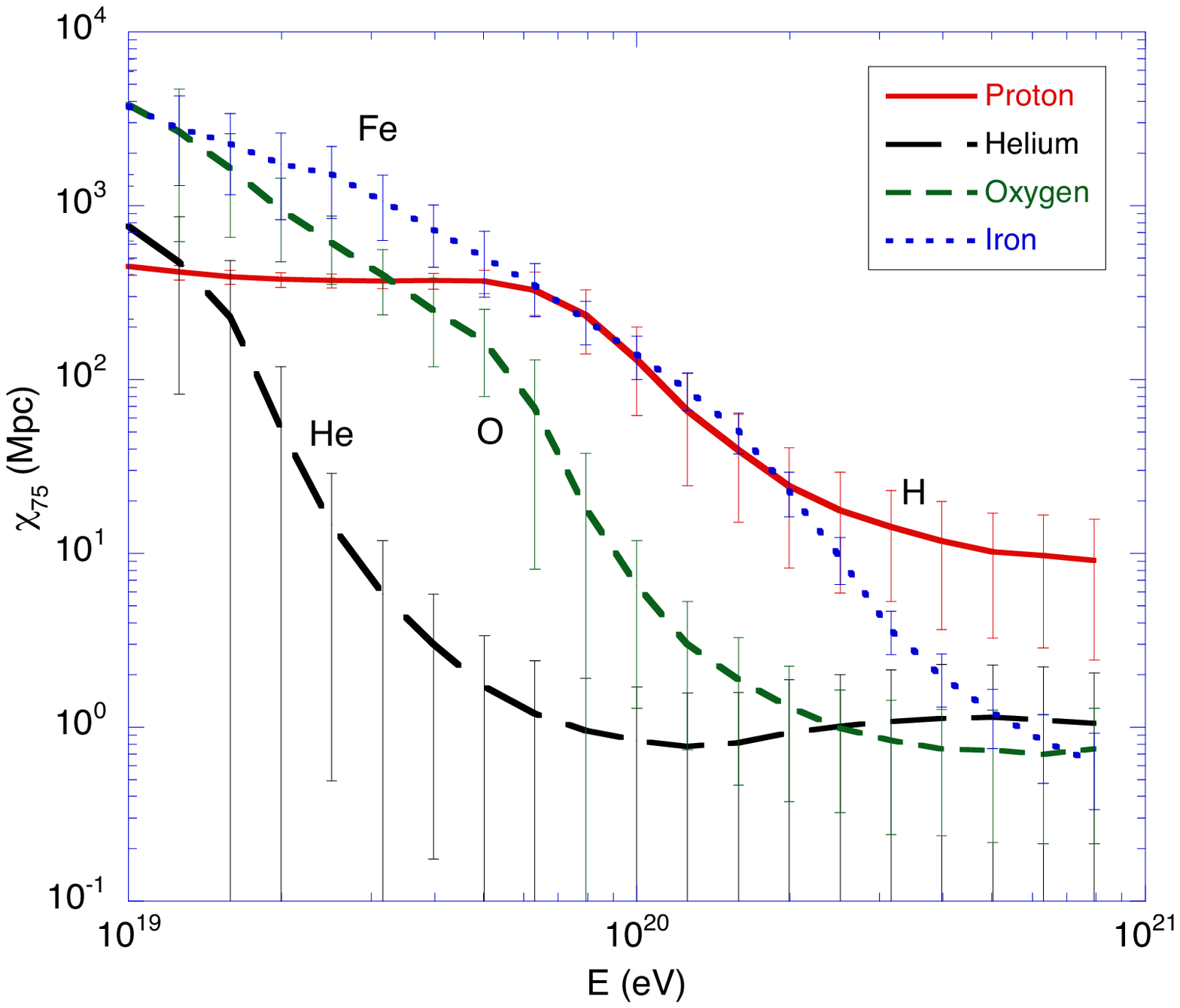}\hfill~
\caption{Left: mean-free-path of $^{56}$Fe nuclei against the various photo-nuclear processes in the CMB and IR background (see text), as a function of $\Gamma$. Right: energy loss length (including pair production) for various nuclei (labeled by their original A) as a function of $E$, with their dispersion.}
\label{fig:MFP}
\end{figure*}

The new GDR cross-sections are discussed in detail in Khan et al. (2005). For the higher-energy processes, we used the parameterisation of Rachen (1996). The QD cross-section is taken as $\sigma_{\mathrm{QD}}(\epsilon,A) = (0.55\,\mathrm{mbarn})\,A^{5/4}(\epsilon/\epsilon_{\mathrm{QD}} - 1)^{3/2}\epsilon/\epsilon_{\mathrm{QD}}^{-3}$, where $\epsilon$ is the photon energy in the nucleus rest frame, $A$ is the nucleon number and $\epsilon_{\mathrm{QD}} \simeq 33.3\,A^{-1/6}$~MeV is the interaction threshold. The number of emitted nucleons is obtained from the experimental branching ratios of Puget et al. (1976). The BR cross-section is modeled as the sum of four Lorentzian representing the four main resonances, $\sigma_{\mathrm{BR}}(\epsilon,A) = A\sum_{i=1}^{4}\sigma_{i}\sigma_{\mathrm{L}}(\epsilon;E_{i},\Gamma_{i})$, with the following parameters (in GeV, GeV and $\mu$barn, respectively): $(E_{i},\Gamma_{i},\sigma_{i}) = (0.34,0.17,351)$, $(0.75,0.50,159)$, $(1.0,0.60,21)$ and $(1.5,0.80,26)$. Finally, the PF cross-section is modeled as $\sigma_{\mathrm{PF}}(\epsilon,A) = (1\,\mu\mathrm{barn})\,A^{0.91}(1 - 2.48\,\exp(-\epsilon/0.8\,\mathrm{GeV}))(69.8\,s^{0.081} + 64.3\,s^{-0.453})$, where $s = 0.88 + \epsilon/(532\,\mathrm{MeV})$.
We also follow Rachen (1996) for the pair production cross-section, including corrections to the formul\ae~of Blumenthal (1970). The attenuation time of a $^{A}_{Z}$X nucleus with Lorentz factor $\Gamma$ is related to that of the proton, $\tau_{\mathrm{p}}$, as: $\tau(Z,A,\gamma) = (A/Z^2)\,\tau_{\mathrm{p}}\times\Phi^{-1}(Z\alpha)$, where $\alpha$ is the fine structure constant and $\Phi(x) = 1 - 0.29x^2 + 0.25x^4 - 0.25x^6$.

The propagation of all the nuclei (distributed in energy according to the source spectrum) is simulated with a Monte-Carlo code taking the above processes into account. We keep track of all secondary nuclei emitted and propagate them in the same way. It is assumed that each ejected fragment (or nucleon) takes away the same energy per nucleon as the parent nucleus, so that it has the same Lorentz factor. In addition to the CMB, we include the IRB as described in Malkan and Stecker (1998): $n(E_{\gamma}) = 1.1\,10^{-4}\,E_{\gamma}^{-2.5}$ for $0.02\,\mathrm{eV}\leq E_{\gamma} \leq 0.8$~eV. The resulting mean free path for a nucleus with Lorentz factor $\Gamma$ is given by:
\begin{equation}
\lambda_{\mathrm{IR}}^{-1} = \frac{1}{2\Gamma^{2}}\int_{\epsilon_{\mathrm{th}}/2\Gamma}^{\epsilon_{\mathrm{max}}}\mathrm{d}\epsilon\frac{n(\epsilon)}{\epsilon^{2}}
\int_{\epsilon_{\mathrm{th}}}^{2\Gamma\epsilon}\epsilon^\prime\sigma(\epsilon^\prime)\mathrm{d}\epsilon^\prime\,,
\label{eq:MFPIR}
\end{equation}
where $\epsilon_{\mathrm{th}}$ is the energy threshold of the process under consideration and $\epsilon_{\mathrm{max}} = 0.8$~eV.

The total mean free path of $^{56}$Fe nuclei against the various photo-nuclear processes in the CMB and IRB is shown in Fig.~\ref{fig:MFP}a as a function of $\Gamma$. The GDR process clearly dominates up to $\Gamma = 10^{10.5}$--$10^{11}$ and is responsible for the main GZK effect for nuclei (sudden reduction of the flux). In Fig.~\ref{fig:MFP}b, we show the length scale of energy losses for protons, He, O and Fe nuclei at the source, defined as the propagation length, $\chi_{75}(E)$, after which a nucleus of initial energy $E$ has lost 25\% of its energy  (including photo-pair production). The error bars indicate the dispersion due to the stochastic propagation, emphasizing the importance of a Monte-Carlo approach. 
As can be seen, the drastic reduction of the propagation horizon, due to the GDR, occurs at different energies for different nuclei and is much more pronounced for heavy nuclei than for protons.
It should be stressed that, while the GZK effect appears stronger for nuclei than for protons, nuclei photoerosion products contribute to the EGCR flux at lower energies and substantially modify the shape of the spectrum. This leads to important changes in the global phenomenology of EGCRs, as we now discuss.

\section{Results and discussion}

\begin{figure*}[ht]
\centering
\hfill\includegraphics[width=7.5cm]{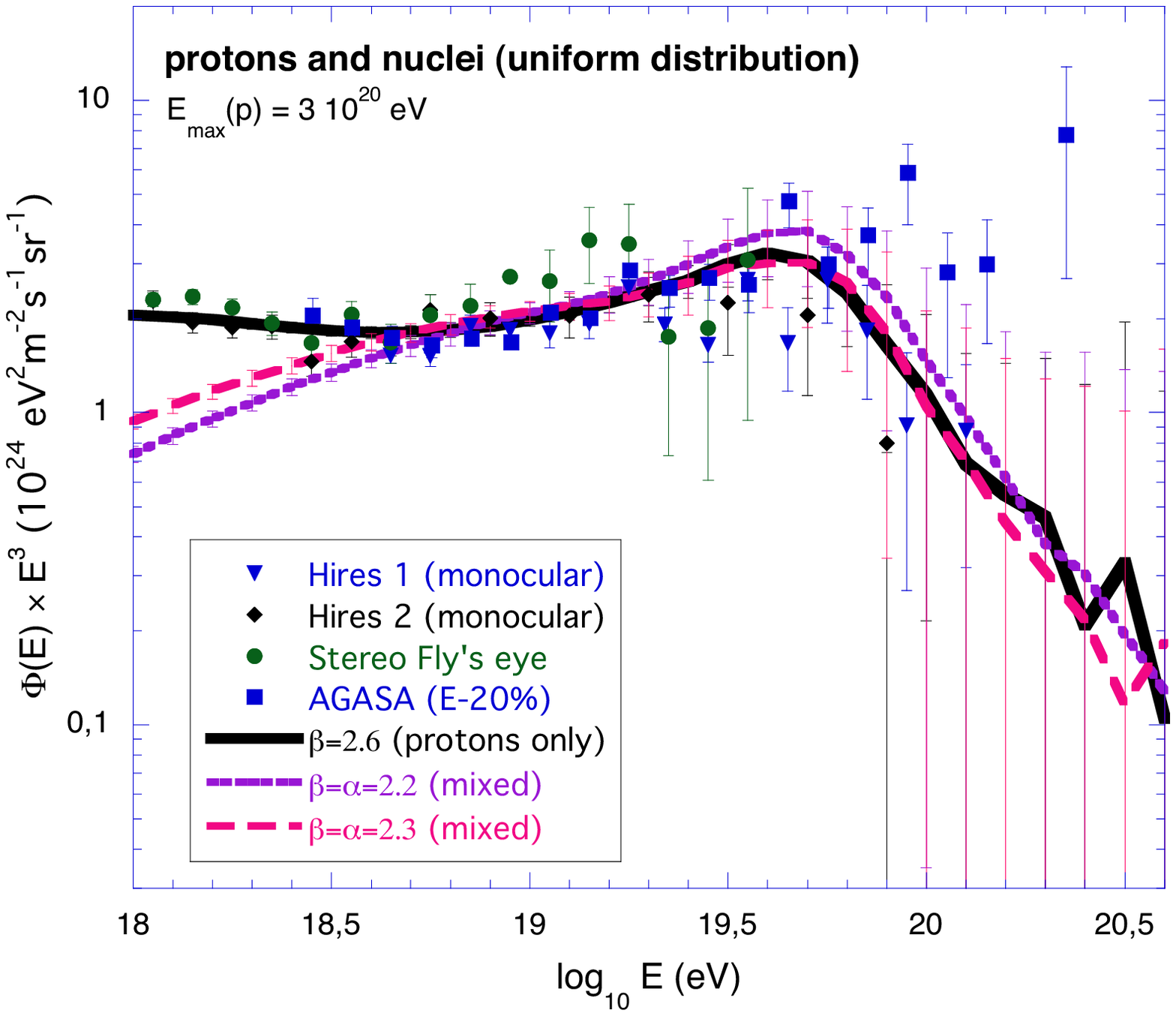}\hfill
\includegraphics[width=7.5cm]{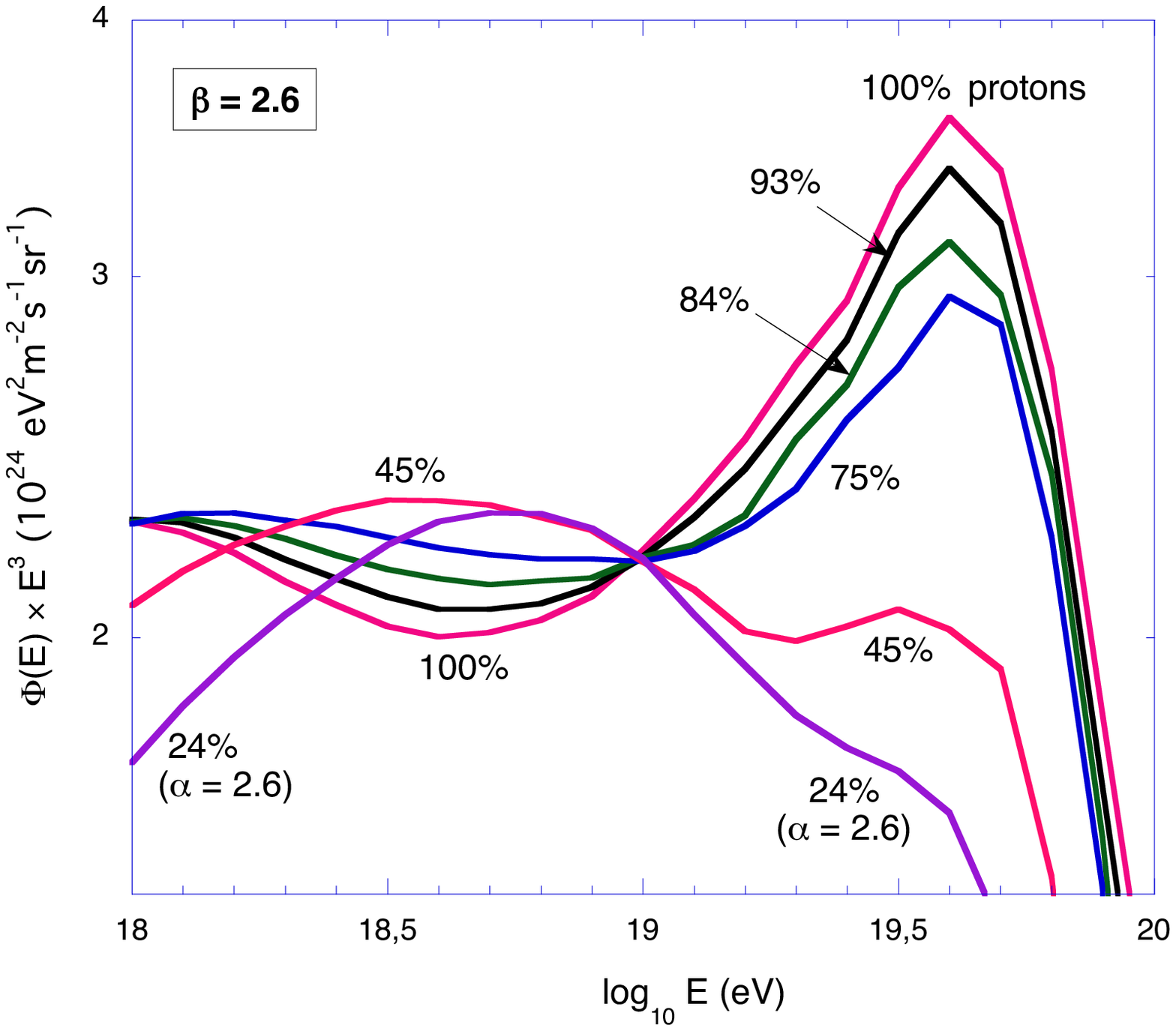}\hfill~
\caption{Left: Propagated spectra for sources with protons only and $\beta = 2.6$, and with mixed compositions and $\beta = \alpha = 2.2$ and 2.3. Data points from the main experiments are indicated, with an arbitrary downward shift of AGASA's energy scale by 20\%. Right: evolution of the spectrum with the fraction of protons at the source, for an $E^{-2.6}$ source spectrum.}
\label{fig:spectra}
\end{figure*}

In Fig.~\ref{fig:spectra}a, we show the resulting propagated spectra obtained with the prescription described above where  $E_{\mathrm{max}}(^{1}_{1}\mathrm{H}) = 3\,10^{20}$~eV.  The error bars represent the 1-$\sigma$ fluctuations of the flux in each energy bin, as expected for a data set with the statistics of the AGASA experiment, i.e., 866 events above $10^{19}$~eV (cf. De Marco et al., 2004). We compare our results with the data from HiRes (Abbusi et al. 2004) and AGASA (Takeda et al. 2003) experiments. In the pure proton case, we find again that the best fit spectrum has $\beta = 2.6$ and a pair production dip, as in previous work. However, if the composition is mixed, the choice of $\alpha = \beta = 2.6$ (corresponding to 24\% of protons at the source above $10^{18}$~eV) provides a very poor fit to the data (cf. Fig.~\ref{fig:spectra}b, lower curve). Figure~\ref{fig:spectra}b shows how the spectrum around the ankle and the GZK bump evolves with the fraction of protons in the source composition (above $10^{18}$~eV). For fractions lower than 75\%, the pair production dip is strongly attenuated and the ankle is actually reversed for fractions lower than $\sim 60$\%. Therefore, the interpretation of the ankle as an e$^{+}$-e$^{-}$ pair production dip requires a very large fraction of protons at the source (see also Berezinsky et al., 2005). In contrast, Fig.~\ref{fig:spectra}a shows two good fits to the highest energy data using mixed composition models with $\alpha = \beta = 2.2$ or 2.3, corresponding to 50\% or 40\% protons at the source, respectively. Note that the observed composition at Earth is much lighter than at the source, because of the massive production of secondary nucleons peaking around $5\,10^{18}$~eV and the successive cut-offs of nuclei with increasing mass (Allard 2004).

As can be seen in Fig.~\ref{fig:spectra}a, the pure proton model with $\beta = 2.6$ provides a good fit to the data down to $\sim 10^{18}$~eV (in a scenario with uniform source distribution and no magnetic field). In such a model, the transition from GCR to EGCR should thus occur at lower energies around the second knee (Berezinsky et al., 2004). However, no acceleration process is known to yield a spectrum as steep as $\beta = 2.6$. Instead, both analytical and numerical studies of particle acceleration in relativistic shocks give injection spectra with  $\beta \simeq 2.2$--2.3  (Bednarz and Ostrowski, 1998; Kirk et al., 2000; Lemoine and Pelletier, 2003). As shown in Fig.~\ref{fig:spectra}a, such spectral indices can provide equally good fits to the high energy data, assuming a realistic source model with a mixed composition $\alpha \simeq \beta$. In such a scenario, the transition from GCRs to EGCRs occurs at the ankle, which thus keeps its ``standard'' interpretation.

In conclusion, our calculations show that the phenomenological interpretation of the CR data at high energy strongly depends on the source composition. We confirm that pure proton sources require a steep spectrum in $\sim E^{-2.6}$ and are able to account for the ankle down to $10^{18}$~eV, interpreted as a pair production dip of the EGCR component. On the other hand, if the EGCRs are accelerated out of the interstellar medium with a source composition roughly similar to that of the GCRs, we find that a harder source spectrum $\sim E^{-2.2}$ gives an equally good fit to the data, allowing an interpretation of the ankle as the transition from GCRs to EGCRs, with the GCR component extending up to $\ga 10^{18}$~eV.

These results are interesting for several reasons. First, a source spectrum with $\beta \simeq$ 2.2--2.3 appears quite plausible from the theoretical point of view (relativistic shock acceleration). Second, a source spectrum $\sim E^{-2.3}$  has also been considered as the best fit to the low-energy data for the GCR component (e.g. Strong and Moskalenko, 2001; Ptuskin, 1997). It is thus particularly interesting to note that assuming a similar composition for both the GCR and EGCR components is precisely what also makes a similar source spectrum possible. In addition, it has been shown in Parizot (2005) that a source power-law index of $\sim 2.3$ is a necessary condition for holistic models, in which the same sources produce the CRs at all energies. The results presented here show that a self-consistent model can be built for CRs with a similar composition and spectrum at all energies provided the GCR/EGCR transition is at the ankle. Finally, the inclusion of nuclei with a standard CR composition at high-energy can avoid the fine-tuning necessary to match the GCR and EGCR components in the second-knee transition models.

\begin{acknowledgements} We thank Maximo Ave and Alan Watson for useful comments. This work was supported in part by the KICP under NSF PHY-0114422, by  NSF  AST-0071235, and  DE-FG0291-ER40606 at the University of Chicago. 
\end{acknowledgements}

\end{document}